\documentstyle [preprint,eqsecnum,aps,epsfig]{revtex}

\def\ll{\langle}
\def\rr{\rangle}

\begin{document}

\vskip 1.0cm

\begin{center}
{\Large\bf High statistics measurement of the underground}
\vskip .4 cm
{\Large\bf muon pair separation at Gran Sasso}

\vskip .6 cm


\bigskip
\begin{center}
{\bf The MACRO Collaboration}\\
\nobreak\bigskip\nobreak
\pretolerance=10000
M.~Ambrosio$^{12}$, 
R.~Antolini$^{7}$, 
C.~Aramo$^{7,n}$,
G.~Auriemma$^{14,a}$, 
A.~Baldini$^{13}$, 
G.~C.~Barbarino$^{12}$, 
B.~C.~Barish$^{4}$, 
G.~Battistoni$^{6,b}$, 
R.~Bellotti$^{1}$, 
C.~Bemporad $^{13}$
% ...
\\
\bigskip
\footnotesize
1. Dipartimento di Fisica dell'Universit\`a di Bari and INFN, 70126 
Bari,  Italy \\
2. Dipartimento di Fisica dell'Universit\`a di Bologna and INFN, 
 40126 Bologna, Italy \\
3. Physics Department, Boston University, Boston, MA 02215, 
USA \\
4. California Institute of Technology, Pasadena, CA 91125, 
USA \\
5. Department of Physics, Drexel University, Philadelphia, 
PA 19104, USA \\
6. Laboratori Nazionali di Frascati dell'INFN, 00044 Frascati (Roma), 
Italy \\
7. Laboratori Nazionali del Gran Sasso dell'INFN, 67010 Assergi 
(L'Aquila),  Italy \\
8. Depts. of Physics and of Astronomy, Indiana University, 
Bloomington, IN 47405, USA \\
9. Dipartimento di Fisica dell'Universit\`a dell'Aquila  and INFN, 
 67100 L'Aquila,  Italy \\
10. Dipartimento di Fisica dell'Universit\`a di Lecce and INFN, 
 73100 Lecce,  Italy \\
11. Department of Physics, University of Michigan, Ann Arbor, 
MI 48109, USA \\	
12. Dipartimento di Fisica dell'Universit\`a di Napoli and INFN, 
 80125 Napoli,  Italy \\	
13. Dipartimento di Fisica dell'Universit\`a di Pisa and INFN, 
56010 Pisa,  Italy \\	
14. Dipartimento di Fisica dell'Universit\`a di Roma ``La Sapienza" and INFN,
 00185 Roma,   Italy \\ 	
15. Physics Department, Texas A\&M University, College Station, 
TX 77843, USA \\	
16. Dipartimento di Fisica Sperimentale dell'Universit\`a di Torino and INFN,
 10125 Torino,  Italy \\	
17. L.P.T.P., Faculty of Sciences, University Mohamed I, B.P. 524 
Oujda, Morocco \\
$a$ Also Universit\`a della Basilicata, 85100 Potenza,  Italy \\
$b$ Also INFN Milano, 20133 Milano, Italy\\
$c$ Also Istituto TESRE/CNR, 40129 Bologna, Italy \\
$d$ Also Scuola Normale Superiore di Pisa, 56010 Pisa, Italy\\
$e$ Also Dipartimento di Fisica, Universit\`a di Roma Tre, Roma, Italy \\
$f$ Also Universit\`a di Trieste and INFN, 34100 Trieste, 
Italy \\
$g$ Also Dipartimento di Energetica, Universit\`a di Roma, 
 00185 Roma,  Italy \\
$h$ Also Institute for Nuclear Research, Russian Academy
of Science, 117312 Moscow, Russia \\
$i$ Also Institute for Space Sciences, 76900 Bucharest, Romania \\
$l$ Swarthmore College, Swarthmore, PA 19081, USA\\
$m$ RPD, PINSTECH, P.O. Nilore, Islamabad, Pakistan \\
$n$ Also INFN Catania, 95129 Catania, Italy\\
$o$ Also Department of Physics, Pennsylvania State University, 
University Park, PA 16801, USA\\
\end{center}

\vskip  4mm
{\bf Abstract}\\
\vskip  2mm
\end{center}

We present a measurement of the underground decoherence function 
using multi-muon events observed in the MACRO detector at 
Gran Sasso at an average depth of 3800~hg/cm~$^2$.
Muon pair separations up to 70 m have been measured, corresponding
to parent mesons with $P_\perp$~$\leq$~1$\div$2~GeV/c.
Improved selection criteria are used to reduce detector effects
mainly in the low distance separation region of muon pairs.
Special care is given to a new unfolding procedure designed to minimize
systematic errors in the numerical algorithm.
The accuracy of the measurement is such that the possible contribution
of rare processes, such as $\mu^\pm +N \rightarrow \mu^\pm + N+\mu^+ + \mu^-$,
can be experimentally studied.

The measured decoherence function is
compared with the predictions of the hadronic
interaction model of the HEMAS Monte Carlo
code. Good agreement is obtained. 
We interpret this agreement to indicate that 
no anomalous $P_\perp$ components
in soft hadron-Nucleus and Nucleus--Nucleus 
collisions are required by the MACRO experimental data. 
Preliminary comparisons
with other Monte Carlo codes point out that 
the uncertainties associated with the hadronic
interaction model may be as large as 20\%, depending on the energy.
MACRO data can be used as a benchmark for future work 
on the discrimination of shower models in the primary energy region 
around and below the knee of the spectrum.

\parskip= 4pt plus 1pt
\textwidth=6.0in
\textheight=8.7in
\oddsidemargin=0.4in
\evensidemargin=0.4in
\headsep=0.1mm
\topmargin=0.001in
\pagestyle{plain}
\setcounter{page}{1}
 
\section {Introduction}

The knowledge of hadronic interaction processes plays a
fundamental role in studies of cosmic rays in the 
VHE--UHE range ($10^{12}$ eV $\leq E \leq 10^{17}$ eV).
In particular, the interpretation of indirect measurements
intended to determine the features of primary cosmic rays, such as 
spectra and composition, depends on the choice of the hadronic 
interaction model adopted in the description of the atmospheric 
shower development.
For instance, muons observed by deep underground experiments are 
the decay products of mesons originating mostly in kinematic regions 
(high rapidity and high $\sqrt{s}$) not completely covered by existing 
collider data.
The problem is particularly important for nucleus-nucleus
interactions for which available data extend only to a few 
hundreds of GeV in the laboratory frame. 
It is therefore crucial to find physical observables which are primarily 
sensitive to the assumed interaction model rather than to the energy spectra 
and chemical composition of primary cosmic rays.

The shape of the 
muon lateral distribution is well-suited for this purpose. In particular it 
allows the study of the transverse structure of hadronic interactions, which
is one of the most relevant sources of uncertainties in the 
models \cite{gaisser}.
In fact, different aspects of the interactions contribute to the
lateral distribution.
We can qualitatively understand this by simple arguments, valid in a
first order approximation. Let us consider a single interaction of
a primary nucleon of total energy $E_0$, producing mesons
of energy $E^{\pi,K}$ with transverse momentum $P_\perp$, at a slant height
$H_{prod}$, which eventually decay into muons. Calling $r$ the separation of 
a high energy muon ({\it i.e.} 
moving along a straight line) from the shower axis, we have: 
\begin{equation}
r \sim \frac{P_\perp}{E^{\pi,K}} H_{prod}.
\end{equation}
In this simplified description we are
neglecting the transverse momentum in the parent decay.
The previous expression can be written in a more instructive
way, considering that at high energy, apart from terms of the order
of $(m_T/E_0)^2$, the longitudinal c.m. variable $x_F$ is approximately
equal to the laboratory energy fraction:
\begin{equation}
r \sim \frac{P_\perp}{x_F^{\pi,K}E_0} H_{prod} \propto
\frac{P_\perp}{x_F^{\pi,K}E_0} \left( \log{\sigma^{inel}_{n-Air}}+const.\right).
\end{equation}
The assumption of an exponential atmosphere 
has been used in the last expression. 
It can be seen how the transverse and longitudinal components of
the interaction, as well as the inclusive and total cross sections,
convolve together (with different weights) to yield the lateral separation. 
The role of $P_\perp$ remains a dominant one in determining the relative
separation of the 
muon component by introducing a loss of collinearity
(``decoherence") with respect to the direction of the
shower axis.

A qualitative extension to the case of nuclear projectiles can be made
within the framework of the superposition model, where each nucleon of the 
projectile
of mass number A is assumed to interact independently with energy $E_0/A$.
Further refinements are needed to account for modifications in the 
$P_\perp$ and $x_F$ distributions deriving
from the nuclear structure of projectile and target, as
will be discussed later. 
A reliable evaluation of the lateral
distribution function can be
obtained only by Monte Carlo methods.

\begin{figure}[thb]
\vskip  2 cm
\begin{center}
\mbox{\epsfysize=9.3cm
      \epsfxsize=13.3cm
      \epsffile{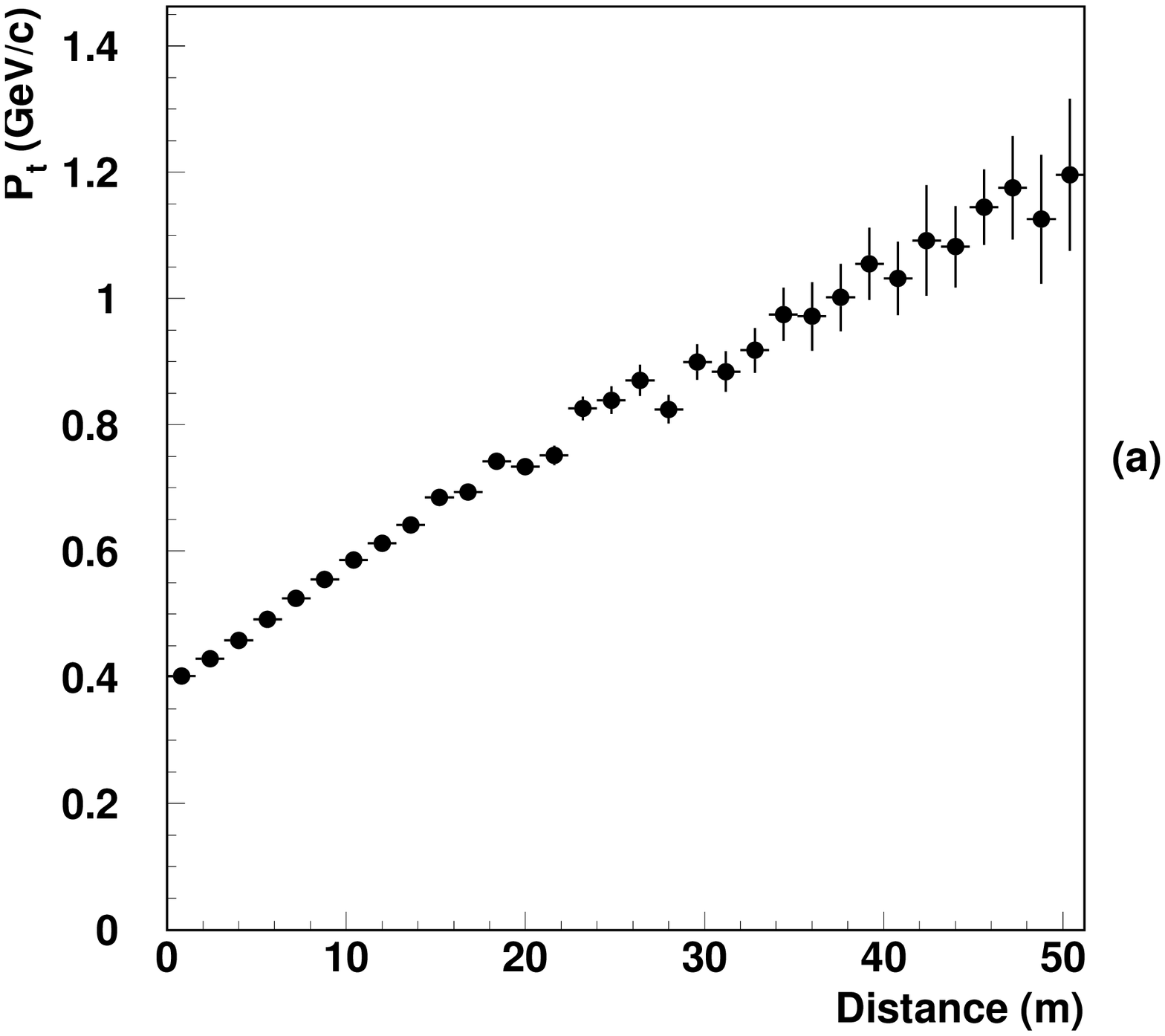}}
\vskip -2 cm
\mbox{\epsfysize=11.1cm
      \epsfxsize=10.8cm
      \epsffile{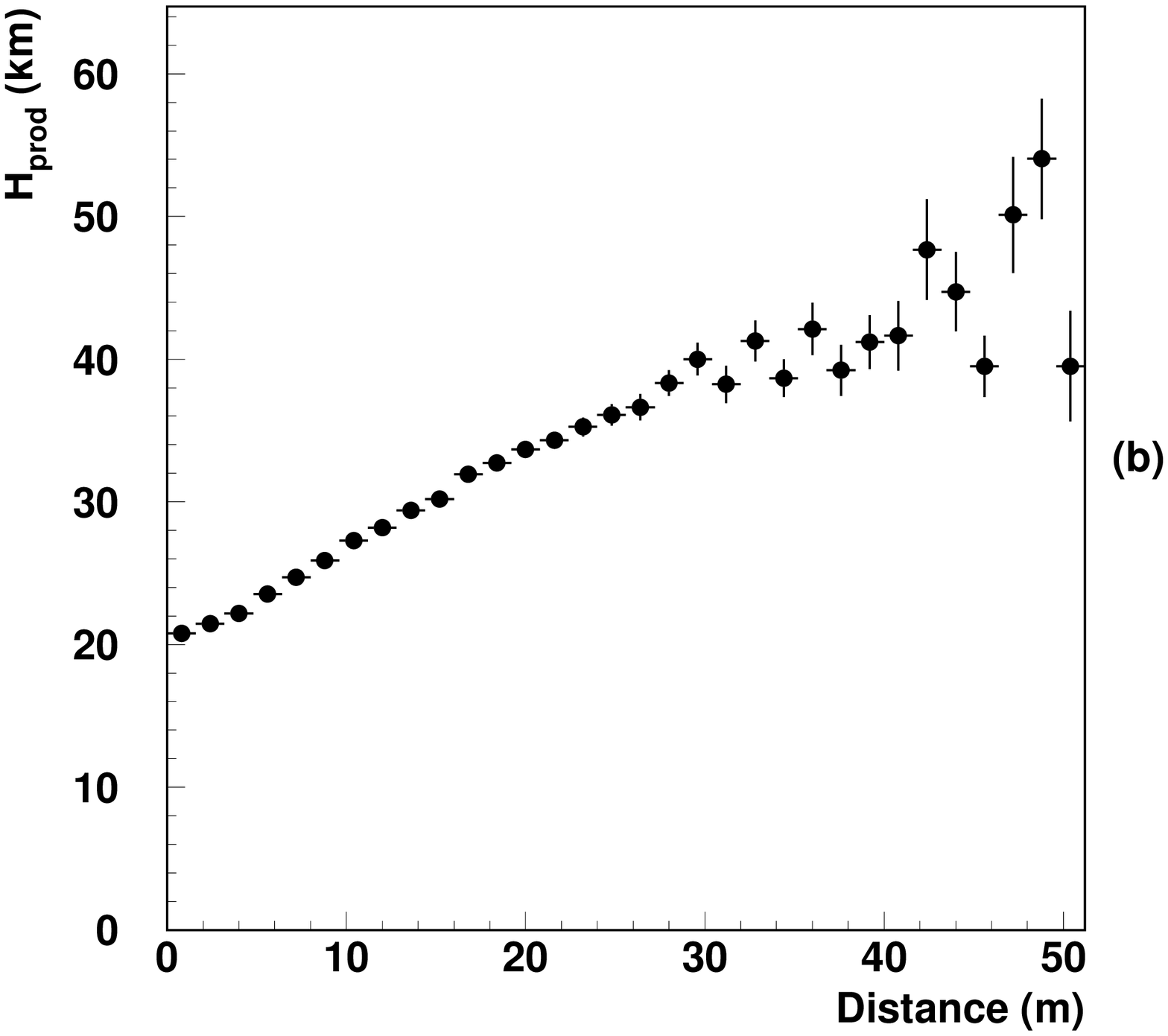}}
\end{center}
\vskip -2 cm
\caption{\em Average separation of underground muon pairs at Gran Sasso 
depth, as a function of $\langle P_\perp \rangle$ of the parent mesons (a) 
and of the slant height in the atmosphere (b). The results are
obtained with the HEMAS Monte Carlo.
\label{f:dvshpt}}
\end{figure}

Deep underground experiments are capable of selecting atmospheric
muons in the TeV range produced in the initial stages of the extensive
air shower (EAS) development. They can perform
a measurement of muon separation which is highly 
correlated to the lateral distribution. Since 
the shower axis position is not usually known, the distribution of
muon pair separation in multimuon events is studied.
Muons associated with the same events, 
coming in general from different parent and shower generations, 
are grouped together.
Furthermore, a wide range of primary energy is 
integrated in the same distribution. It is generally assumed, and
supported by many simulations, that the shape of this
distribution is only slightly affected by the mass composition of 
primaries \cite{sokolsky}, 
thus preserving the sensitivity to the interaction features.
As an example, in Fig. \ref{f:dvshpt} we show the dependence of
the average pair separation, as detected at the depth of the underground
Gran Sasso laboratory, with respect to the 
$\langle P_\perp \rangle$ of the parent mesons and to  
their production slant height in the atmosphere.
These have been calculated by means of the
HEMAS Monte Carlo code \cite{hemas} for a mixed primary 
composition \cite{macro_newcomp2}.
This code employs an interaction model based on the results of the experiments
at hadron colliders.

The decoherence function as measured in an underground
experiment is also affected by multiple scattering in the rock and, 
to some extent, geomagnetic deflection.

For a detector with geometrical acceptance
$A(\theta,\phi)$, for zenith and azimuthal angles $\theta$ and $\phi$,
respectively, we define the decoherence function as the distribution of the 
distance between muon pairs in a bundle:
\begin{equation}
\frac{dN}{dD}=\frac{1}{\Omega T}\int{\frac{1}{A(\theta,\phi)}
\frac{d^{2}N(D,\theta,\phi)}{dD d\Omega}d\Omega},
\label{eq:deco}
\end{equation}
where $N(D,\theta,\phi)$ is the number of muon pairs with a separation $D$
in the direction ($\theta$,$\phi$),
$\Omega$ is the total solid angle covered by the apparatus
and $T$ is the total exposure time of the experiment.
A muon bundle event of multiplicity $N_\mu$ will contribute with
a number of independent pairs $N = N_\mu(N_\mu-1)/2$.

In principle, a decoherence study can be performed without a single
large area detector, and in early attempts the muon lateral separation 
was studied via coincidences between two separate movable detectors 
\cite{utah}. 
The advantage offered by a single large area detector
is the ability to study the features inherent in 
the {\it same} multi-muon event, such as higher 
order moments of the decoherence distribution \cite{clusters}.

The large area MACRO detector \cite{macro_detect}
has horizontal surface area of $\sim 1000$~m$^2$ at an average depth of
3800 hg/cm$^2$ of standard rock ($E_\mu\geq~$1.3 TeV) and  
is naturally suited for
this kind of measurement. An analysis of the muon decoherence has
already been performed 
\cite{macro_deco,calgary,macro_newcomp1,macro_newcomp2}.
The bulk of multiple muon events in MACRO corresponds to a selection of 
primary energies between a few tens to a few thousands of TeV/nucleon.
Hadronic interactions and shower development in the atmosphere
were simulated with the previously noted HEMAS code.
In particular, a weak dependence on primary mass
composition was confirmed for two extreme cases:
the ``heavy'' and ``light'' composition models \cite{adelaide}.
The MACRO analysis was designed to unfold the true muon decoherence 
function from the measured one by properly considering
the geometrical containment and track resolution efficiencies.
This procedure permits a direct comparison between measurements
performed by different detectors at the same depth, and, more
importantly, whenever new Monte Carlo simulations are available, allows
a fast comparison between predictions and data without the need to
reproduce all the details of detector response.

The first attempt, 
obtained while the detector was still under construction, 
and therefore with a limited size,
was presented in \cite{macro_deco}.
The same analysis, with a larger sample based on the full lower detector,
was extended in \cite{calgary}.
With respect to the HEMAS Monte Carlo expectations, these results 
indicated a possible excess in the observed distribution at large separations.
In Ref. \cite{macro_newcomp2} we presented the decoherence distribution 
without the unfolding procedure; the claimed excesses 
were not confirmed.
In order to reach more definitive conclusions, a more careful analysis of 
the systematics associated with the unfolding procedure
was considered necessary.
A detailed discussion of this item will be addressed in Section~4.

A more careful discussion of the Monte Carlo simulation is also necessary.
The bulk of the muon bundles collected by MACRO are low multiplicity events, 
coming from parent mesons in the far forward region of UHE
interactions, not easily accessible with collider experiments.
This requires an extrapolation to the highest energies and rapidity 
regions, introducing possible systematic uncertainties.
For instance, some doubts have been raised \cite{gaisser} concerning 
the treatment of meson $P_\perp$ in HEMAS.
In the HEMAS hadronic interaction code, secondary particle $P_\perp$ 
depends upon three different contributions:
\begin{itemize}
\item $\langle P_\perp \rangle$ increases with energy, 
as required by collider data in the central region;
\item $\langle P_\perp \rangle$ increases in p--Nucleus and 
Nucleus--Nucleus interactions, relative to that for pp collisions, 
according to the ``Cronin effect'' \cite{cronin};
\item $\langle P_\perp \rangle$ varies with $x_{F}$, according
to the so called ``seagull effect'' \cite{seagull}.
\end{itemize}
The sum of these effects yields some doubt about a possible 
overestimate of $P_\perp$ for energetic secondary particles,
an hypothesis recently restated in \cite{soudan2}.
It is therefore crucial to perform a high precision test of the transverse
structure of this model, since it affects the calculation of
containment probability for multiple muon events and, consequently,
the analysis of primary cosmic ray 
composition \cite{macro_newcomp1,macro_newcomp2}. 

In this paper, a new analysis of the unfolded decoherence function 
is presented, performed with improved methods up to 70 m.
The present work enlarges and completes the data analysis presented in 
\cite{macro_newcomp1,macro_newcomp2}.
Preliminary results of this unfolding procedure \cite{durban} showed an 
improved agreement between experimental data and Monte Carlo predictions.

Particular attention is paid to the small-separation ($D\leq 1$ m)
region of the decoherence curve, in which processes such as
muon-induced hadron production can produce a background to the high
energy muon analysis.
At the energies involved in the present analysis ($E_{\mu}\geq$~1~TeV),
moreover, muon-induced muon pair
production in the rock overburden could yield an excess of events with
small separation, as suggested in \cite{olga}.
 This process is usually neglected in 
Monte Carlo models commonly adopted for high energy muon transport 
\cite{hemas,fluka,geant,lipari_stanev}.

Section~2 is devoted to the description of the detector and 
of data analysis, with a focus on new event selection criteria.
In Section~3 the features of the Monte Carlo simulation are presented
together with the comparison between experimental and simulated data
in the MACRO detector, 
while Section~4 is dedicated to the unfolding procedure.
A comparative discussion of the features of different hadronic interaction 
models is summarized in Section~5.
In Section~6, the problem existing in the first bins of the decoherence 
distribution is presented in detail, testing new hypotheses on its origin.
Conclusions follow in Section~7.

\section {Detector description and data analysis}

The MACRO detector \cite{macro_detect}, located in hall B of the 
Gran Sasso Laboratory, is a large area detector equipped with streamer 
tube chambers, liquid scintillation counters and nuclear track detectors 
arranged in a modular structure of six ``supermodules''.
Each of these is 12 m$\times$12 m$\times$9 m in size
and consists of a 4.8 m high lower level and a 4.2 m upper ``attico''.
In this paper only data from the lower level of the apparatus are 
included; therefore only the lower detector will be described further.

Tracking is performed by means of limited streamer tubes, which are distributed 
in ten horizontal planes separated by $\sim$ 60 g cm$^{-2}$ of
CaCO$_3$ (limestone rock) absorber, and in six planes along each vertical wall.
The streamer tubes have a square cross section of 3$\times$3 cm$^2$, 
and are 12 m long.
From each plane two coordinates are provided, the wire (perpendicular 
to the long detector dimension) and strip views. The latter employs 
3 cm wide aluminum strips at 26.5$^\circ$ to the wire view.
The average efficiencies of the streamer tube and strip systems 
were 94.9\% and 88.2\% respectively, in the period of this analysis.

The spatial resolution achieved with this configuration depends
on the granularity of the projective views. The average
width of a cluster, defined as a group of contiguous muon "hits,"
is 4.5 cm and 8.96 cm for the wire and strip views, respectively.
Muon track recognition is performed by an algorithm which requires 
a minimum number of aligned clusters (usually 4) through which
a straight line is fit. The differences between the cluster centers
and the fit determine
a spatial resolution of $\sigma_{W}$=1.1 cm for the wire view and 
$\sigma_{S}$=1.6~cm for the strip view.
These resolutions correspond to an intrinsic angular resolution
of $0.2^{0}$ for tracks crossing ten horizontal planes.

In reconstructing the best bundle configuration, the tracking package 
flags track pairs as parallel, overlapping, or independent and not 
parallel. This is achieved in two steps, in each projective view:
\begin{itemize}
\item two tracks are defined as parallel if their slopes coincide within
  2 $\sigma$ or if their angular separation is less than $3^{0}$
  ($6^{0}$ if the tracks contain clusters whose widths exceed 30~cm).
  Otherwise, the track pair is flagged as independent and not
  parallel if its distance separation is larger than 100 cm. 
\item tracks at short relative distance are labelled 
  as overlapping if their intercepts with the detector bottom level
  coincide within 3.2 $\sigma$ (2 $\sigma$ if their angular separation 
  is $<1.5^{0}$).
\end{itemize}
The routine chooses the most likely bundle as the set having the largest
number of parallel tracks and the largest number of points per track.
Subsequently, tracks flagged as not parallel are considered in order to
include fake muon tracks originated primarily by hadrons or $\delta$-rays 
in the surrounding rock or inside the detector. 
A two-track separation of the order of 5 cm is achieved on each projective
view. However, this capability can be substantially worsened in case
of very large, but rare, catastrophic energy losses of muons in the detector.

Only tracks with a unique association in the two views can be 
reconstructed in three dimensional space.
At this level, pattern recognition is used to require a complete
matching between tracks belonging to different projective views.
This is automatically achieved when two tracks pass through
separate detector modules. 
When they are in the same module, matching of hit wires and strips 
on the same detector plane is accomplished by taking advantage of 
the stereo angle of the strips with respect to the wires. In some 
cases the track pattern correspondence between the two views is also used.
The possibility to analyse muon decoherence in three dimensional space
is important to have an unbiased decoherence distribution.
However, the unambiguous association of muon tracks from the two projective 
views cannot be accomplished for high multiplicity events because, in 
events characterized by a high muon density, the tracking algorithm is 
not able to resolve the real muon pattern without ambiguities, especially 
when tracks are superimposed.
In Ref. \cite{macro_newcomp1,macro_newcomp2} we presented the muon 
decoherence function in the wire view alone, which allowed 
the extension of the analysis to higher multiplicities.

We have analyzed about $3.4 \cdot 10^{5}$ events, corresponding 
to a 7732 hr live time for the lower part of the apparatus. 
These events were submitted to the following selection criteria:
\begin{enumerate}
\item Zenith angle smaller than $60^0$. This choice is dictated 
by our limited knowledge of the Gran Sasso topographical map for 
high zenith angles. 
Moreover, we cannot disregard the atmosphere's
curvature for larger zenith angles, which at present our current
simulation models do not include.
\item Fewer than 45 streamer tube hits out of track.
This selection is designed to eliminate possible 
misleading track reconstruction in events produced by noise in the
streamer tube system and/or electromagnetic interactions in or near 
the apparatus.
\item Track pairs must survive the parallelism cut.
This rejects hadrons from photonuclear interactions close to
the detector, as well as tracks reconstructed from electromagnetic 
interactions which survived the previous cut.
\end{enumerate}

The last cut is not completely efficient in rejecting muon
tracks originating from local particle production because 
the angle between these tracks may fall within the limits 
imposed by the parallelism cut.
These limits cannot be further reduced since the average
angular divergence due to multiple muon scattering in the rock 
overburden is about $1^{0}$ at the MACRO depth.
This is a crucial point, since these events could
contaminate the decoherence curve in the low separation region
and are not present in the simulated data because of the excessive CPU
time required to follow individual secondary particles.
A similar effect could be produced by single muon tracks with
large clusters, which may be reconstructed as a di-muon event 
by the tracking algorithm.

In order to reduce these effects, a further selection was 
applied. We computed, for each muon track in the wire view, the ratio 
$R$ between the number of streamer tube planes hit by the muon to the 
number of planes expected to be hit considering the track direction. 
Only tracks with $R \geq 0.75$ were accepted.
The application of this cut (hereafter cut C4) in the wire view 
alone is a good compromise between the rejection capability of the algorithm 
and the loss of events due to the unavoidable inefficiency of the streamer 
tube system. We found that in the wire projective view the probability 
to reject a muon track due to contiguous, inefficient planes is 2.0\%.

\begin{figure}[thb]
\begin{center}
\mbox{\epsfysize=12cm 
      \epsffile{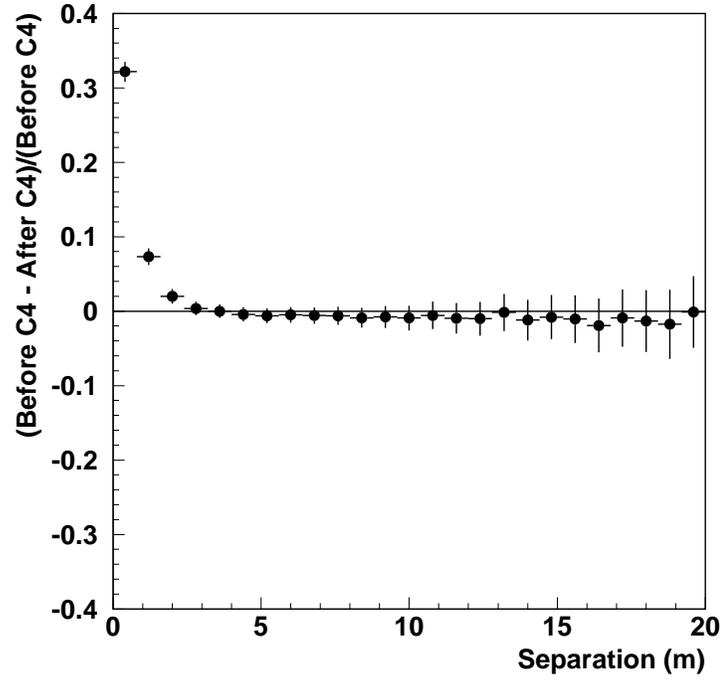}}
\end{center}
\caption{\em The change in the experimental decoherence function 
induced by cut C4. The data indicate the fractional deviation 
between the experimental decoherence function before and after the
application of the cut. 
\label{f:r75}}
\end{figure}

To show the effects produced by cut C4, we present in 
Fig. \ref{f:r75} the fractional differences between the experimental
decoherence curve before and after its application.
As expected, the new cut affects only the first bins of the distribution.

\begin{figure}[thb]
\begin{center}
\mbox{\epsfysize=12cm 
      \epsffile{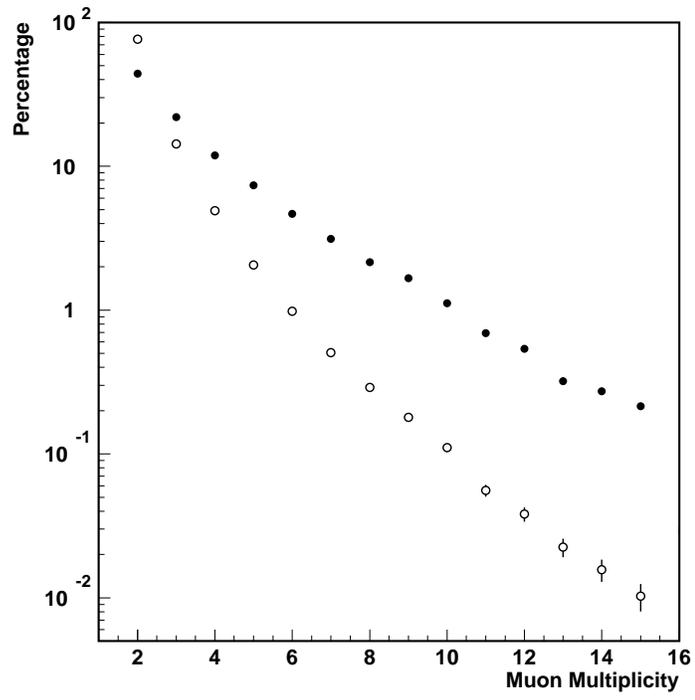}}
\end{center}
\caption{\em Percentage of reconstructed real events (white points) 
and unambiguously associated muon pairs (black points) as a function 
of event multiplicity. 
\label{f:stat}}
\end{figure}

To test the ability of cut C4 to reject hadronic tracks,
we used FLUKA \cite{fluka} to simulate 3028 hr of live time 
in which muons were accompanied by hadronic products of photonuclear
interactions in the 10 m of rock surrounding the detector.
We found that the parallelism cut alone provides a rejection efficiency
of about 54.6\% of the pair sample, while the addition of cut C4 enhances the 
rejection to 95.9\%. The effect of hadron contamination, furthermore,
is very small, contributing less than 1\% in the overall muon pair sample.
This estimate, together with the plot of Fig. \ref{f:r75} and the results
of a visual scan, suggest that
the main track sample rejected by cut C4 is made of large cluster tracks.
After the overall application of these cuts, the number of 
surviving unambiguously associated muon pair tracks is 355,795.
In Fig. \ref{f:stat} the percentage of the reconstructed 
events as a function of muon multiplicity is shown (open circles).
In the same figure, the percentage of the unambiguously associated muon pairs 
as a function of the multiplicity is also reported (black circles).
Due to detector effects, the number of associated pairs $N'_{pair}$ in an
event of multiplicity $N_\mu$ is generally smaller than the maximum number of 
independent pairs $N_{pair} = N_\mu(N_\mu-1)/2$.
This reduction becomes greater for high multiplicities, for obvious 
reasons of track shadowing.
In any case, we still find that the weight of high multiplicity events remains
dominant in the decoherence distribution. 
In order to reduce this effect and to reduce the possible dependence on
 primary composition, we have assigned a weight $1/N'_{pair}$ to each
 entry of the separation distribution. This prescription, followed
 also for simulated data, has been already applied 
in most of the previous analyses 
performed by MACRO.
Moreover, we emphasize that the focus of this analysis is centered
on the shape of the distribution; the absolute rate
of pairs as a function of their separation is neglected.

\section {Monte Carlo Simulation}

The Monte Carlo chain of programs used in the simulation consists
of an event generator,
capable of following the development of the hadronic shower in the atmosphere
and the muon transport code in the rock overburden, and a detector
simulation package.
We have used, as in all
previous relevant analyses of muon events in MACRO 
\cite{macro_newcomp1,macro_newcomp2}, the HEMAS code \cite{hemas}
as an interaction model and shower simulator. 
Nuclear projectiles are handled by interfacing HEMAS with the 
``semi-superposition'' model of the NUCLIB library \cite{nuclib}.
The final relevant piece of simulation is the 
three-dimensional description of muon transport in the rock.
A comparison of the performance of different transport codes,
reported in Ref. \cite{propmu}, showed that 
the original package contained in the HEMAS code was too simplified,
leading, for instance, to an underestimated muon
survival probability at TeV energies. In order to verify 
possible  systematics affecting the decoherence distribution,
we repeated the Monte Carlo production, interfacing the
more refined PROPMU code \cite{lipari_stanev} to HEMAS.
We have verified that, at least to first approximation, no 
changes in the shape of the decoherence function are noticeable
between the two different simulation samples.
For this reason, the sum of the two different Monte Carlo productions
will be used in the following.

The map of Gran Sasso overburden as a function of
direction and the description of its chemical composition
are reported in Ref. \cite{verth}.
The detector simulation is based on the CERN package GEANT \cite{geant}.
The folding of simulated events with the detector simulation 
is performed according to a variance reduction method \cite{battistoni} 
to minimize statistical losses and reducing possible systematic errors.

We generated $3.6\cdot 10^{8}$ primary interactions in the total energy range
3$\div$10$^5$ TeV, assuming the ``MACRO-fit'' primary mass composition model 
\cite{macro_newcomp1,macro_newcomp2}, in which five mass groups 
(p, He, CNO, Mg and Fe) are considered. 
Simulated data are produced with the same format as real data and
are processed using the same analysis tools.
After the application of the same cuts as for real data, a sample of about
$7.0 \cdot 10^{5}$ muon pairs survived, corresponding to about 645 days 
of MACRO live time.

\begin{figure}[thb]
\begin{center}
\mbox{\epsfysize=12cm 
      \epsffile{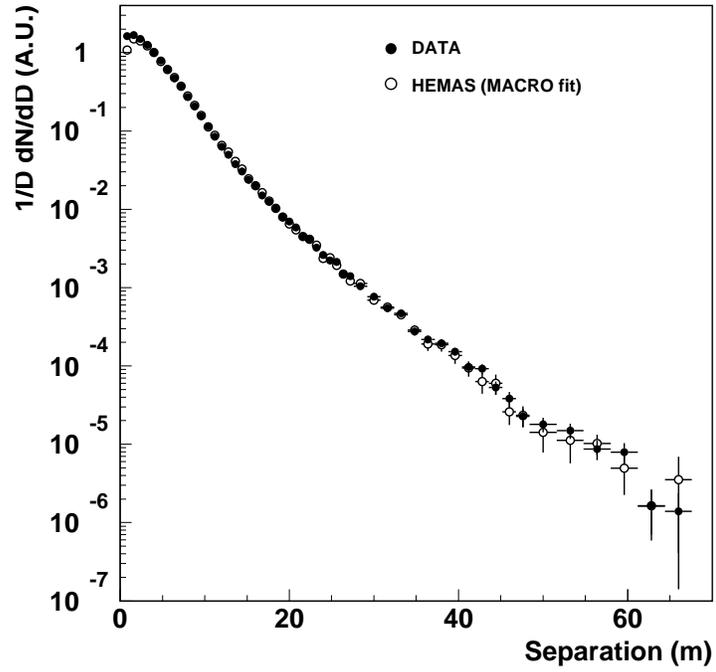}}
\end{center}
\caption{\em Experimental (black points) and simulated (white points)
decoherence function, normalized 
to the peak of the $dN/dD$ distribution. The second to last points
of the two distributions coincide. 
\label{f:folded}}
\end{figure}

In Fig. \ref{f:folded} the comparison between the 
experimental and simulated decoherence curve inside the detector is shown. 
Curves are normalized to the peak of the $dN/dD$ distribution.
The remarkable consistency of the two curves demonstrates the HEMAS code 
capability to reproduce the observed data up to a maximum distance of 70 m.
The bump in the experimental distribution around 40 m is due to the 
detector acceptance and is visible also in the simulated data, thus
confirming the accuracy of our detector simulation.
We also notice that, despite the application of cut C4,
there is a non-negligible discrepancy between the experimental and 
simulated data in the first two bins of the distribution
of ($34\pm 2$)\% and ($10\pm 1$)\%, respectively. 
Such a discrepancy is not predicted by any model, since at short
distances, apart from detector effects, the shape of decoherence 
distribution is dictated by
the solid angle scaling: $dN/dD^2|_{D\rightarrow 0} \sim const.$, 
while the
relevant properties of the interactions under investigation manifest
themselves in the shape at large distances.
The origin of this discrepancy 
 will be discussed in detail
in the last section of this paper, where other sources of contamination
in the real data sample will be taken into account. 

\section {Unfolding Procedure}

The agreement of the Monte Carlo and data shown in Fig. \ref{f:folded} 
proves that the simulation is consistent with observation and that
the detector structure is well reproduced. A detector-independent
analysis is required in order to subtract the geometric effects
peculiar to MACRO, and allows a more direct comparison with other
analyses and/or hadronic interaction models. This is accomplished
by a correction method, built with the help of the Monte Carlo simulation,
to unfold the ``true''
decoherence function from the measured one in which
geometrical containment and track reconstruction efficiencies are considered.

\begin{figure}[thb]
\begin{center}
\mbox{\epsfysize=12cm 
      \epsffile{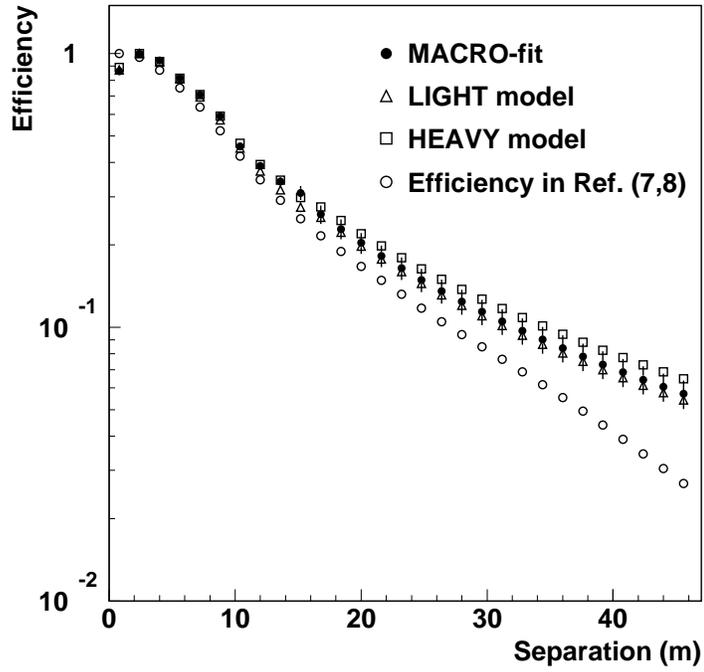}}
\end{center}
\caption{\em Comparison of the unfolding efficiencies as a function 
of pair separation for different composition models. The curves are 
normalized to the peak value. For comparison, we include the efficiency 
evaluated with the method used in previous analyses (white points). 
\label{f:eff}}
\end{figure}

In the previous decoherence studies \cite{macro_deco,calgary}, 
the unfolding procedure was 
based on the evaluation of the detection efficiency for 
di-muon events generated by the Monte Carlo with a given angle and
separation. Although this method is composition independent and 
allowed us to determine the detector acceptance with
high statistical accuracy, it introduces systematic effects that have
so far been neglected.
In particular, in a multi-muon event it may happen
that in a given projective view and in a particular
geometrical configuration one muon track is ``shadowed'' by another. 
To avoid this effect, we adopt the following new unfolding method:
the efficiency evaluation is performed considering the whole sample
of events generated with their multiplicities. 
For a given bin of ($D$,$\theta$,$\phi$), where $D$ is the muon pair
separation and ($\theta$,$\phi$) is the arrival direction of the event,
we calculate the ratio
\begin{equation}
\epsilon(D,\theta,\phi) = 
\frac {N^{in}(D,\theta,\phi)}{N^{out}(D,\theta,\phi)}
\label{eq:epsilon}
\end{equation}
between the number of pairs surviving the
selection cuts $N^{in}$
and the number of pairs inserted in the detector simulator $N^{out}$.
In principle, this choice of $\epsilon$ could be dependent on
the primary mass composition model, since 
for a fixed distance $D$ the efficiency (\ref{eq:epsilon}) is dependent 
on the muon density and hence on its multiplicity,
which in turn is correlated with the average atomic mass
$\langle A \rangle$ of the primary.
To check the systematic uncertainty related to this possibility, 
we evaluated the decoherence distributions obtained by unfolding the 
experimental data assuming the ``heavy'' and ``light'' composition 
models. Fig. \ref{f:eff} shows the relative comparison of the shape 
of the unfolding efficiencies as a function of pair separation, 
integrated in $(cos\theta,\phi)$ after the normalization to the peak value.
In the same plot we present the unfolding efficiency calculated
with the method used in Ref. \cite{macro_deco,calgary}. 
Considering the effect of the normalization, we observe that
this method tends to overestimate the efficiencies in the low 
distance range, a consequence of the shadowing effect as explained 
in Section 2.

The unfolded decoherence is given by
\begin{equation}
\left(\frac{dN}{dD}\right)_{unf}=\sum_{(\theta,\phi)} 
\frac {N^{exp}(D,\theta,\phi)}
{\epsilon(D,\theta,\phi)},
\label{eq:unfolding}
\end{equation}
where $N^{exp}(D,\theta,\phi)$ is the number of muon pairs
detected with a separation $D$.
In practice, we used 50 windows in 
$(cos\theta,\phi)$ space (5 and 10 equal intervals for 
$\cos\theta$ and $\phi$, respectively).

The ability to evaluate the integral 
\ref{eq:unfolding} for separate and independent 
windows constitutes a powerful check of the 
systematics related to the decoherence dependence 
on the variables $(cos\theta,\phi)$.
Unfortunately this is not possible for $r$
larger than 45 m, due to insufficient statistics.
In that case the observables $N^{in},N^{out}$ and $N^{exp}$ are integrated 
over $(cos\theta,\phi)$. 
We verified that the systematic error introduced by that choice 
is smaller than the present statistical error in that distance range.

\begin{figure}[thb]
\begin{center}
\mbox{\epsfysize=12cm 
      \epsffile{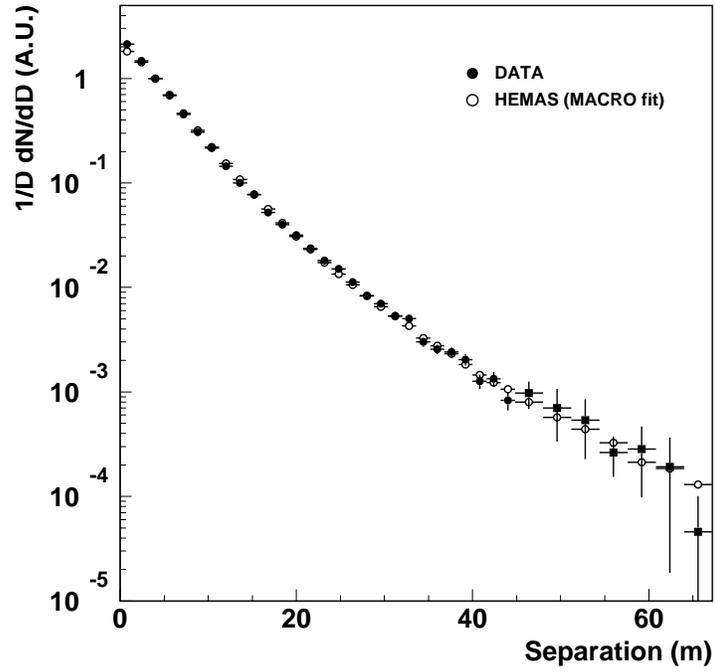}}
\end{center}
\caption{\em Unfolded experimental decoherence distribution compared 
with the infinite-detector Monte Carlo expectation, computed 
with the HEMAS interaction code and the MACRO-fit primary 
composition model. Black squares represent data above 45 m (integral
form unfolding).
\label{f:unfolded}}
\end{figure}

Finally, unfolded experimental data obtained with the MACRO-fit
model are directly compared with the Monte Carlo simulation 
(Fig. \ref{f:unfolded}).
The two curves are in good agreement although the disparity in the 
first bin of the distribution remains unresolved (see Section VI).
The experimental values of the $dN/dD$ distributions, normalized to the peak
value, are reported in Table \ref{t:m1}.

\section {Uncertainties of the hadronic interaction model}

The present work, as are others from MACRO, is extensively based on 
the HEMAS code.
This was explicitly designed to provide a fast tool for
production of high energy muons ($E_\mu>$500 GeV).
However, as mentioned before, the interaction model of HEMAS is based
on parametrizations of existing accelerator data and therefore is subject 
to the same risks of all this class of simulation codes.
In particular, important correlations might be lost, or wrong, or
the necessary extrapolations
required by the specific kinematic regions of cosmic ray physics 
could yield unrealistic results.
This remains a central problem of cosmic ray physics.
For this reason in the last few years general interest has grown 
in ``physically inspired" simulations.
These are based upon theoretical and phenomenological models
like QCD and the Dual Parton Model \cite{dpm}, capable of properly 
constraining the predictions where data do not exist, 
without the introduction of a large number of free parameters.
It is worthwhile to mention the attempt to merge the DPMJET 
model \cite{dpmjet} into the shower simulation of HEMAS \cite{hemasdpm}, and
the interface of the CORSIKA shower code \cite{corsika} with
different models, like HDPM (the original interaction model of
CORSIKA), VENUS \cite{venus}, QGSJET \cite{qgsjet},
SIBYLL \cite{sibyll}, and the afore mentioned DPMJET.
A review of general results obtained using CORSIKA with those models
has been provided by the Karlsruhe group \cite{karls}. 
A common feature of all these models is the more or less direct reference
to the Regge-Gribov theories \cite{gribov} 
for the soft contribution (low $P_\perp$).
It must be stressed that such a phenomenological framework, by its nature,
provides only predictions for the longitudinal properties of the interaction.
The transverse structure leading to the specific $P_\perp$ distribution is not
constrained by the theory, except for the higher $P_\perp$ phenomena, where
perturbative QCD can be used (this is of small relevance in the primary 
energy region addressed by the MACRO data).
Once again, the model builders have to be guided mostly by experimental data,
introducing {\it a-priori} functional forms along with their additional
required parameters. 
Some of the quoted models introduce proper recipes for the continuity
between the soft and perturbative QCD regimes, and also specific
nuclear phenomena like the Cronin effect mentioned above 
(see for instance \cite{dpmjet}).
In practice, the only  possibility to evaluate a
systematic uncertainty associated with the simulation model 
(at least those concerning the transverse structure of the showers)
is to compare the predictions 
from all these models, HEMAS included.
For this purpose, since the Karlsruhe report \cite{karls} did not address this
point, we have performed test runs with some of the models interfaced to 
CORSIKA, to which PROPMU \cite{lipari_stanev} has also been interfaced by us 
for muon transport in the rock overburden.
A full simulation with all the other codes was outside our present 
capability, so we limited ourselves to comparisons at 
a few fixed primary energies, and at fixed primary angles of 
30$^\circ$ in zenith and 190$^\circ$ in azimuth.
These correspond to an average rock overburden of $\sim$ 3200 hg/cm$^2$.
In Tables \ref{tabc1} we show this comparison for a few
representative average quantities for 3 different primary proton energies.
We have considered the average depth of the first interaction 
$X$, $\langle P_\perp \rangle$ for pions coming from the first interaction, 
the average production slant height 
$H_{\mu}$ of muons surviving underground (the decay height of their parent
mesons\footnote{CORSIKA does not allow direct access to the production
height of parent mesons, which would be more interesting for our purposes}.), 
the average distance of the muons from shower axis $\langle R \rangle$ 
and the average underground decoherence $\langle D \rangle$. 
Before discussing the results, it is important to remark that as far single
interactions are concerned, all the models considered give a $P_\perp$
distribution following, with good approximation, the typical power 
law suggested by accelerator data.
This is  $\propto 1/(P_\perp + P_0)^\alpha$, although with somewhat
different parameters for different models. 
Older models, like those predicting a simple exponential
distribution for $P_\perp$, cannot reproduce the muon lateral distribution 
observed in MACRO data \cite{macro_deco}.

In the energy range of 100-1000 TeV, to which most of MACRO data
belong, the resulting differences in the average muon separation do not exceed
20\%. These discrepancies seem to reduce at higher energy, while they appear
much larger at few tens of TeV.
DPMJET is probably the only model predicting a higher average separation than
HEMAS. 
A precise analysis of the reasons leading to the differences among models
is complicated. However, we note that HEMAS gives in general higher values of
average $P_\perp$ than the other models. 
The only exception is indeed DPMJET, which, as mentioned before, pays
particular attention to the reproduction of nuclear effects 
affecting the transverse momentum, as measured in heavy ion 
experiments \cite{heavy}. On the other hand, the effect of
this large $P_\perp$ on the lateral distribution of muons is moderated 
in HEMAS by a deeper shower penetration 
(the inelastic cross section is  based on Ref. \cite{durand}); 
in general HEMAS exhibits a somewhat smaller height of meson production.
\par\noindent 
Similar features in the comparison of models are also obtained 
for nuclear projectiles.
It is therefore conceivable that, for the same primary spectrum and composition,
not all the models considered could reproduce the MACRO decoherence 
curve.
Thus the best fit for spectrum and composition as derived from the analysis of
muon multiplicity distribution in MACRO will also probably differ according 
to the model. 

At least in part, the decoherence analysis can disentangle 
different ranges of longitudinal components of the interaction
from the transverse ones, if this is performed in different 
zenith angle and rock depth windows.
In fact, larger zenith angles correspond (on average) 
to larger muon production slant heights.
This is a geometrical effect due to the greater distance from the
primary interaction point to the detector for large zenith angle 
and consequently to the greater spreading of the muon bundle before 
reaching the apparatus. 
Larger rock depths select higher energy muons and consequently higher 
average energy of their parent mesons.
The average separation decreases with the rock depth since,
qualitatively, the longitudinal momentum 
$\langle P_\parallel \rangle$ increases linearly with energy while 
$\langle P_\perp \rangle$ increases only logarithmically.
The overall result of increasing rock depth is the production 
of final states in a narrower forward cone, decreasing the muon 
pair average separation observed at the detector level. 

\begin{figure}[thb]
\begin{center}

\mbox{\epsfysize=20cm
      \epsfxsize=15cm
      \epsffile{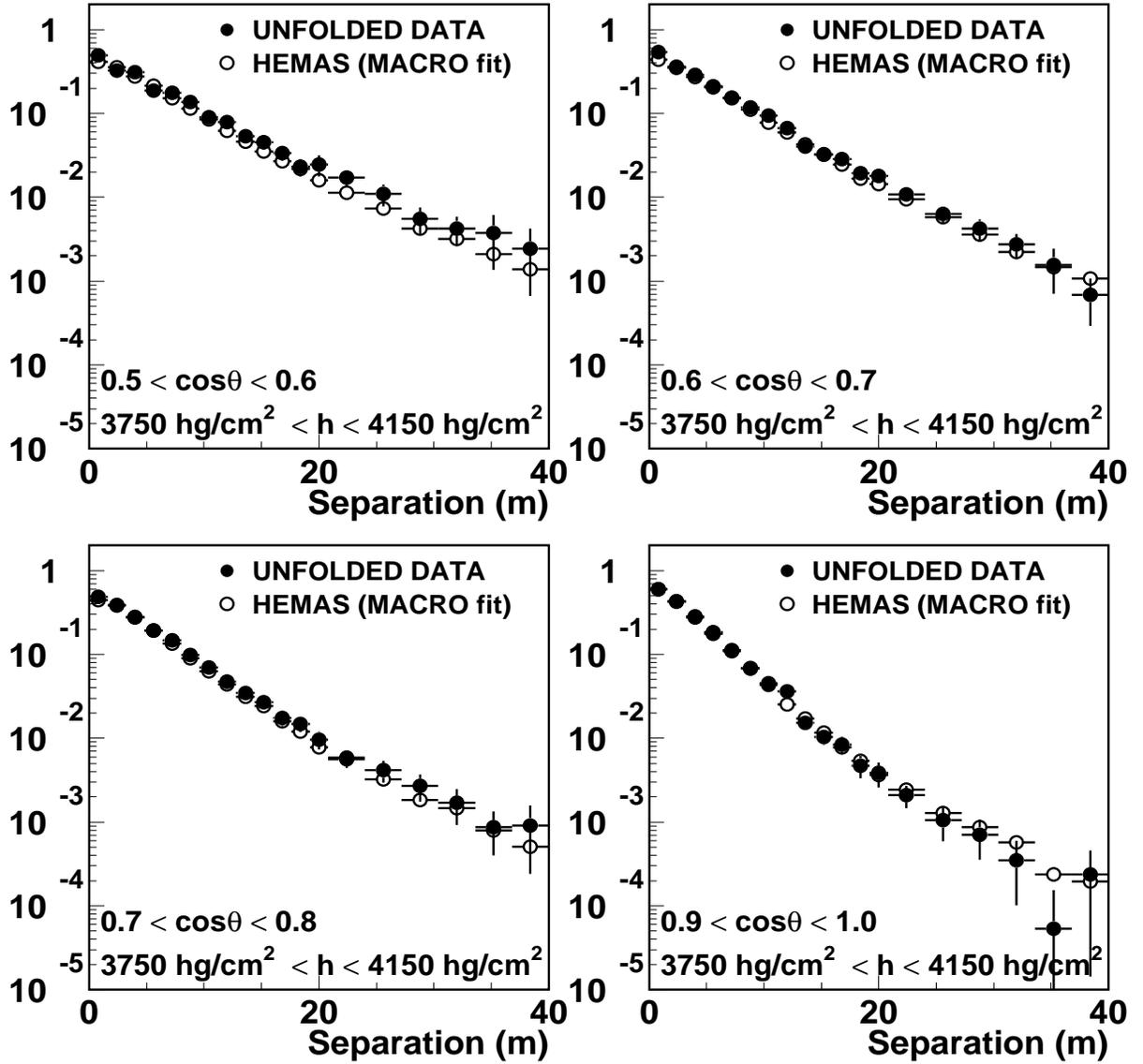}}
\end{center}
\caption{\em Unfolded decoherence functions compared with Monte Carlo 
simulations for different $cos\theta$ windows. 
The vertical scale is in arbitrary units. 
\label{f:win_theta}}
\end{figure}

\begin{figure}[thb]
\begin{center}

\mbox{\epsfysize=20cm
      \epsfxsize=15cm
      \epsffile{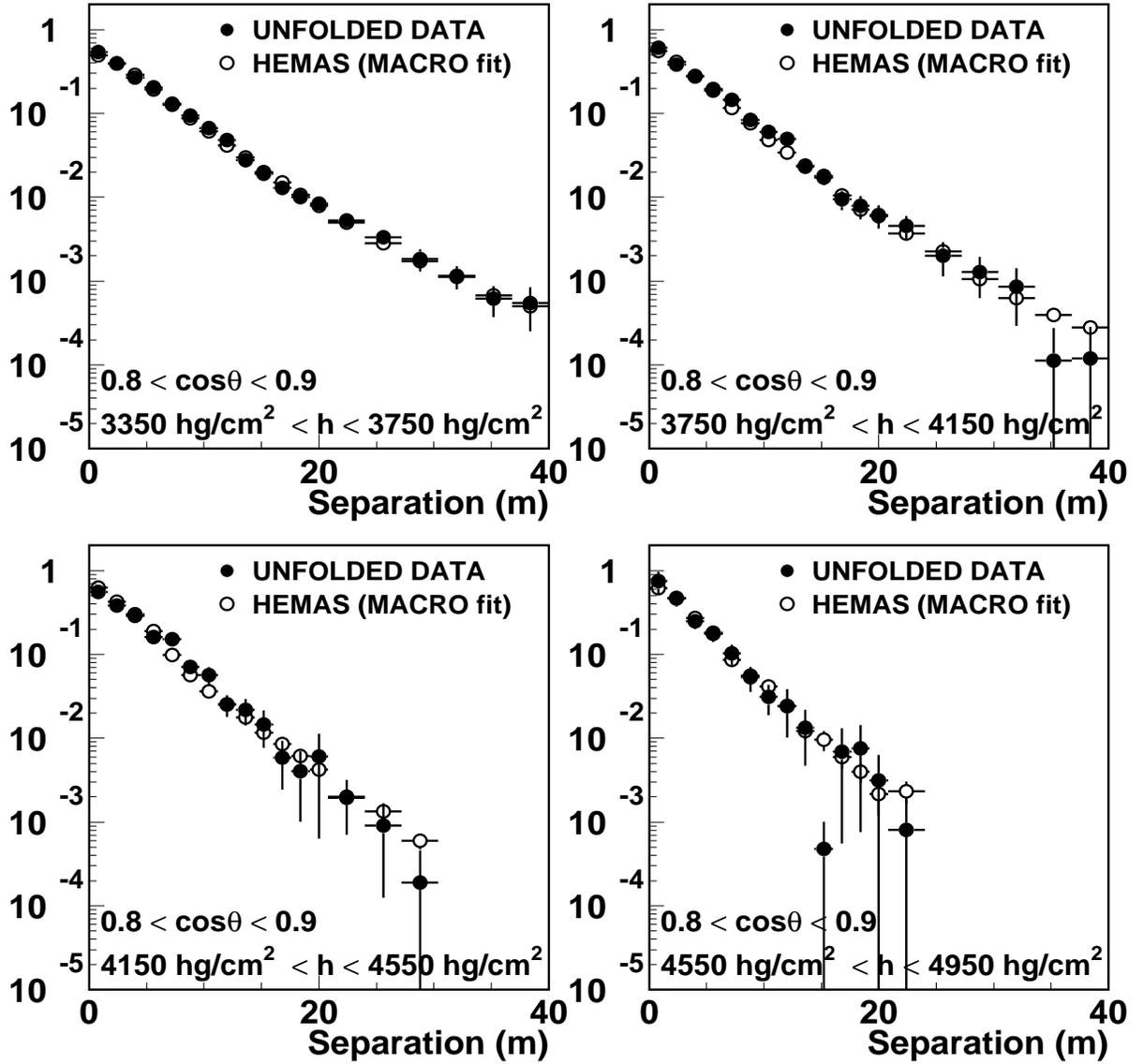}}
\end{center}
\caption{\em Unfolded decoherence functions compared with Monte Carlo 
simulations for different rock depth windows. 
\label{f:win_rock}}
\end{figure}

In Fig. \ref{f:win_theta} and \ref{f:win_rock} the 
unfolded decoherence function is compared to the HEMAS prediction
for different zenith and rock depth intervals.
In Table \ref{t:tab0}, the average separation $\langle D \rangle$ is 
reported as a function of $cos\theta$ and rock depth for
fixed rock depth and zenith, respectively. 
In the same table we report the average values of slant
 height of first interaction $\langle X \rangle$, 
muon production slant height $\langle H_\mu \rangle$,
energy $\langle E_p \rangle$ and transverse momentum 
$\langle P_\perp \rangle$ of the parent mesons, 
as obtained from the HEMAS Monte Carlo in the same windows.

The agreement between the results and the Monte Carlo 
in separate variable intervals reinforces our confidence in
the capability of HEMAS to reproduce the significant
features of shower development. 
This also allows us to exclude the existence of significant 
systematic errors related to this analysis.

\section {The contribution of the 
$\mu^\pm +N \rightarrow \mu^\pm + N+\mu^+ + \mu^-$
process at small distances}

The capability of the MACRO detector to resolve very closely spaced tracks
permits the extension of the decoherence analysis to a distance 
region hardly studied in the past. The mismatch 
between experimental and simulated data in this region ($D\leq$ 160 cm) 
has been emphasized earlier in our discussion. In Section 2 a solution was 
attempted, permitting us to discard, with high efficiency, 
those tracks originating from secondary particle production.
However, Fig.~\ref{f:folded} and Fig.~\ref{f:unfolded} show 
that other sources of contamination in the first bin of the 
decoherence function are responsible for the discrepancy.

The process of muon pair production by muons in the rock, 
$\mu^\pm +N \rightarrow \mu^\pm + N+\mu^+ + \mu^-$,
is a natural candidate. As pointed out in \cite{olga}, at the
typical muon energy involved in underground analyses 
($E_{\mu} \sim$ 1 TeV)
and for very large energy transfer,
the cross section for this process is non-negligible with respect to 
e$^+$e$^-$ pair production.
An analytic expression for the muon pair production cross section is
given in \cite{olga,kelner}. 
In order to test the hypothesis, such a cross section 
has been included in the muon transport code PROPMU.
Assuming a muon flux with energy spectrum $E^{-3.7}$ 
and minimum muon energy $E_{\mu}^{min}$ = 1.2 TeV at the surface,
and considering the actual mountain profile, we generated a sample of 
$10^{7}$ muons corresponding to 3666 h of live time.
About $\sim 3.0 \cdot 10^{6}$ muons survived to the MACRO level,
5360 of which were generated by muon pair production processes.
The average separation of these muon pairs is $(128\pm1)$ cm, and
their average residual energies are $(657\pm14)$~GeV and $(145\pm3)$~GeV,
respectively, for the main muon and the secondary muon samples.
We propagated the muons surviving to the MACRO level 
through the GEANT simulation and we applied the same cuts specified 
in Section~2. Finally, the number of events was normalized
to the live time of real data.

In Table \ref{t:tab1} we report the number of weighted muon pairs
in the first bins of the experimental and simulated decoherence 
distributions (in the form $dN/dD$).
The effect of standard cuts, of cut C4, and of the subtraction 
of the muon pair production process are shown in order. 
In each case, we indicated in percentage the bin populations with 
respect to the peak of the distribution and the discrepancy with
respect to the Monte Carlo predictions.

\begin{figure}[thb]
\begin{center}
\mbox{\epsfysize=12cm 
      \epsffile{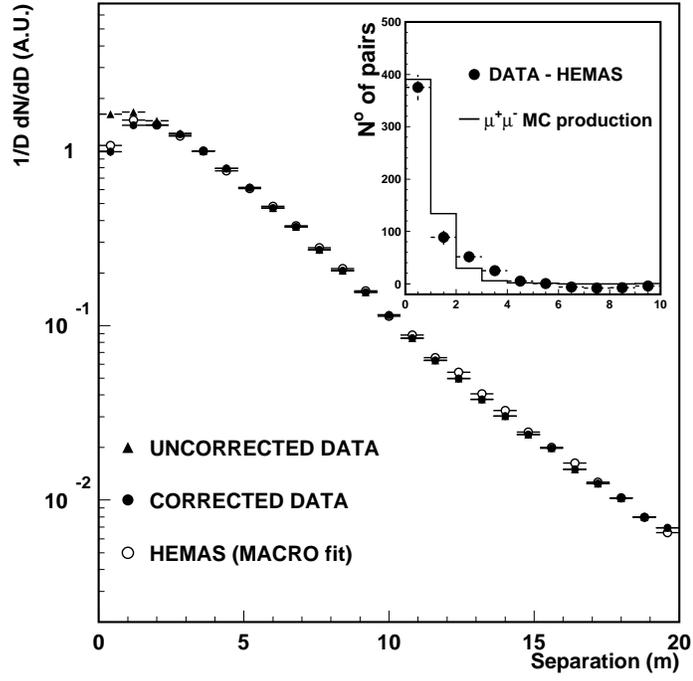}}
\end{center}
\caption{\em The low distance region of the experimental 
decoherence function, before and after the subtraction of the secondary 
muon sample, and comparison with the Monte Carlo simulation. 
The inset shows the
distribution of relative distance for muon pairs in excess
of the data after the subtraction of the HEMAS prediction, as compared
to the expectation from simulated muon pair production in the rock.
\label{f:deco_clean}}
\end{figure}

In Fig. \ref{f:deco_clean} we compare the simulated decoherence curve
with the data corrected for the muon pair production effect.
Despite the approximation introduced in our test, 
it seems that the proposed muon pair production process
can account for most of the observed discrepancy in the low distance range.
This is also shown in the inset of Fig. \ref{f:deco_clean} where the
distribution of relative distance for the muon pairs in excess
of the data (after subtraction of HEMAS prediction) is compared
to the expectation from simulated muon pair production.

An excess at small pair separation is also predicted in exotic processes,
like multi-W production by AGN $\nu$'s, as suggested in\cite{morris}.
However, according to this reference, muons from W$\rightarrow$$\mu$+$\nu$
decay have an average energy of $\simeq$80 TeV. These muons would survive
underground with a residual energy much higher than that of standard muons,
producing local catastrophic interaction in the detector, making difficult
their identification as a pair.
On the contrary, the explanation proposed here is based on a pure QED 
process that does not require any additional physics.

\section {Discussion and conclusions}

We have obtained an improved experimental underground decoherence function
using high energy muons ($E_{\mu}> 1.3~TeV$) 
up to a maximum distance of about 70 m.
It is hard to conceive in the near future a 
large area underground experiment capable of improving the 
sensitivity reached in this decoherence study.

A new unfolding of the experimental distribution confirms the results 
obtained with the analysis within the detector. 

The ability to resolve closely spaced muon tracks allows an investigation
of the decoherence function at small separations.
Apart from the negligible contamination of hadro-production by muons
(which will be the subject of a future work), we found that a relevant
contribution is made by the process 
$\mu^\pm +N \rightarrow \mu^\pm + N+\mu^+ + \mu^-$. The inclusion 
of this interaction 
in the simulation reproduces, in both a qualitative and
a quantitative way, the experimental data.

The agreement of the overall distribution shape for experimental 
and simulated data from HEMAS is excellent. 
The possible excess at high muon separations  suggested 
in the previous, preliminary, analyses \cite{macro_deco,calgary}
was due to an imperfect unfolding procedure, and
is now excluded.
These results both in the integrated distribution and in those from
separate intervals of zenith and rock depth, 
shows that HEMAS gives a reasonable account of the
cascade development and that it is not necessary to introduce any anomalous
$P_\perp$ production in the Monte Carlo to reproduce these data. 
However, the other interaction models considered for comparison, 
while reproducing similar behaviour, in general give 
different combinations of transverse momentum and production height.
Discrimination among the different models may be
possible only after a complete simulation and analysis of MACRO data with each
of the codes.
Therefore the present work, representing a final
 data reduction and analysis, provides a valuable benchmark for future
 analysis dedicated to the investigation
of the properties of high energy interactions and to the evaluation of 
different
shower models in the primary energy region spanning from a few tens to a 
few thousands TeV/nucleon. 
The detector independent analysis described here will make this task easier.



\begin{table}[p]
\begin{center}
\begin{tabular}{c c c}
D (cm)   &   $\frac{dN}{dD}$ & Error \\
\tableline
    80      &   0.4254           &  0.1034 $10^{-1}$ \\
   240      &   0.8855           &  0.1591 $10^{-1}$ \\
   400      &   1.0000           &  0.1738 $10^{-1}$ \\
   560      &   0.9591           &  0.1719 $10^{-1}$ \\
   720      &   0.8148           &  0.1554 $10^{-1}$ \\
   880      &   0.6730           &  0.1419 $10^{-1}$ \\
  1040      &   0.5595           &  0.1350 $10^{-1}$ \\
  1200      &   0.4341           &  0.1216 $10^{-1}$ \\
  1360      &   0.3410           &  0.1118 $10^{-1}$ \\ 
  1520      &   0.2939           &  0.1144 $10^{-1}$ \\
  1680      &   0.2198           &  0.4950 $10^{-2}$ \\
  1840      &   0.1828           &  0.4626 $10^{-2}$ \\
  2000      &   0.1578           &  0.4476 $10^{-2}$ \\
  2160      &   0.1283           &  0.4134 $10^{-2}$ \\
  2320      &   0.1047           &  0.3906 $10^{-2}$ \\
  2480      &   0.9348 $10^{-1}$ &  0.3881 $10^{-2}$ \\
  2640      &   0.7438 $10^{-1}$ &  0.3599 $10^{-2}$ \\
  2800      &   0.5847 $10^{-1}$ &  0.3211 $10^{-2}$ \\
  2960      &   0.5168 $10^{-1}$ &  0.3238 $10^{-2}$ \\
  3120      &   0.4173 $10^{-1}$ &  0.2994 $10^{-2}$ \\
  3280      &   0.4113 $10^{-1}$ &  0.3125 $10^{-2}$ \\
  3440      &   0.2582 $10^{-1}$ &  0.2585 $10^{-2}$ \\
  3600      &   0.2315 $10^{-1}$ &  0.2444 $10^{-2}$ \\
  3760      &   0.2260 $10^{-1}$ &  0.2522 $10^{-2}$ \\
  3920      &   0.2004 $10^{-1}$ &  0.2501 $10^{-2}$ \\ 
  4080      &   0.1289 $10^{-1}$ &  0.2023 $10^{-2}$ \\
  4240      &   0.1419 $10^{-1}$ &  0.2319 $10^{-2}$ \\ 
  4400      &   0.9105 $10^{-2}$ &  0.1775 $10^{-2}$ \\
  4560      &   0.6776 $10^{-2}$ &  0.1597 $10^{-2}$ \\
  4720      &   0.3080 $10^{-2}$ &  0.1035 $10^{-2}$ \\
  4880      &   0.2990 $10^{-2}$ &  0.1121 $10^{-2}$ \\
  4640      &   0.1128 $10^{-1}$ &  0.3340 $10^{-2}$ \\
  4960      &   0.8697 $10^{-2}$ &  0.4549 $10^{-2}$ \\
  5280      &   0.7108 $10^{-2}$ &  0.4098 $10^{-2}$ \\
  5600      &   0.3689 $10^{-2}$ &  0.1534 $10^{-2}$ \\
  5920      &   0.4190 $10^{-2}$ &  0.2742 $10^{-2}$ \\
  6240      &   0.2991 $10^{-2}$ &  0.2703 $10^{-2}$ \\
  6560      &   0.7539 $10^{-3}$ &  0.9004 $10^{-3}$ \\
\end{tabular}
\end{center}
\caption{\em Tabulation of the the unfolded decoherence distribution
as measured by MACRO. The data points are normalized to the point of maximum.
\label{t:m1}}
\end{table}

\begin{table}[p]
\begin{center}
p--Air, 20 TeV
\vskip 0.2cm
\begin{tabular}{l c c c c c}
Code    & $\ll X_{first}\rr $  & $\ll P_\perp\rr $ $\pi^\pm$  & $\ll H_{\mu}\rr $ & 
$\ll R\rr $ & $\ll D\rr $ \\ 
                & (g/cm$^2$) & (GeV/c) & (km)  & (m)    &  (m)  \\
\tableline
HEMAS           &  51.4   &  0.40  & 24.1 &  7.9  & 12.7 \\
CORSIKA/DPMJET  &  44.4   &  0.42  & 25.6 & 10.1  & 13.9 \\
CORSIKA/QGSJET  &  45.7   &  0.39  & 24.3 &  7.3  & 10.0 \\
CORSIKA/VENUS   &  48.3   &  0.35  & 24.5 &  7.4  &  8.3 \\
CORSIKA/SIBYLL  &  50.9   &  0.37  & 23.5 &  7.2  & 11.5 \\
\end{tabular}
\vskip 0.5cm
p--Air, 200 TeV
\vskip 0.2cm
\begin{tabular}{l c c c c c}
Code    & $\ll X_{first}\rr $  & $\ll P_\perp\rr $ $\pi^\pm$  & $\ll H_{\mu}\rr $ & 
$\ll R\rr $ & $\ll D\rr $ \\ 
                & (g/cm$^2$) & (GeV/c) & (km)  & (m)    &  (m)  \\
\tableline
HEMAS           &  56.1   &  0.44  & 20.6 &  5.3 & 8.0 \\
CORSIKA/DPMJET  &  53.9   &  0.43  & 21.7 &  6.2 & 8.8 \\
CORSIKA/QGSJET  &  52.8   &  0.41  & 21.4 &  5.5 & 7.8 \\
CORSIKA/VENUS   &  60.2   &  0.36  & 20.9 &  5.3 & 7.5 \\
CORSIKA/SIBYLL  &  55.2   &  0.41  & 20.2 &  5.2 & 7.3 \\
\end{tabular}
\vskip 0.5cm
p--Air, 2000 TeV
\vskip 0.2cm
\begin{tabular}{l c c c c c}
Code    & $\ll X_{first}\rr $  & $\ll P_\perp\rr $ $\pi^\pm$  & $\ll H_{\mu}\rr $ & 
$\ll R\rr $ & $\ll D\rr $ \\ 
                & (g/cm$^2$) & (GeV/c) & (km)  & (m)    &  (m)  \\
\tableline
HEMAS           &  63.0   &  0.50  & 16.3 &  4.1  & 6.0 \\
CORSIKA/DPMJET  &  60.0   &  0.42  & 18.5 &  4.9  & 6.4 \\
CORSIKA/QGSJET  &  63.1   &  0.44  & 17.7 &  4.2  & 5.6 \\
CORSIKA/VENUS   &  66.7   &  0.36  & 16.8 &  4.1  & 5.3 \\
CORSIKA/SIBYLL  &  60.3   &  0.44  & 17.0 &  4.4  & 5.6 \\
\end{tabular}
\vskip 0.4cm
\caption{\em Comparison of a few relevant quantities concerning the
lateral distribution of underground muons at the depth of 3200
hg/cm$^2$, from proton primaries at 20, 200 and 2000 TeV, 
30$^\circ$ zenith angle. 
The statistical errors are smaller than the last reported digit.}
\label{tabc1}
\end{center}
\end{table}

\begin{table}[p] 
\begin{center}
3750$<$h$<$4150 ($hg/cm^{2}$)
\vskip 0.2cm
\begin{tabular}{c l c c c c c}
			 &
			 &
   .5$<$cos$\theta$$<$.6 &
   .6$<$cos$\theta$$<$.7 &
   .7$<$cos$\theta$$<$.8 &
   .8$<$cos$\theta$$<$.9 &
   .9$<$cos$\theta$$<$1. \\
\tableline
 EXP & $\ll D\rr $ (m)           & $13.2\pm 2.3$&$11.4\pm 2.2$&$10.3\pm 2.2$&$8.5\pm 1.9$&$7.5\pm1.9$\\
\tableline
     & $\ll D\rr $ (m)           & $12.8\pm 1.4$&$12.0\pm 1.3$&$10.1\pm 1.2$&$8.8\pm 1.2$&$7.8\pm1.1$\\
     & $\ll X\rr $ (km)          & $65.9\pm0.2$ & $57.0\pm 0.2$ & $51.2\pm 0.3$ & $42.5\pm 0.4$ & $37.0\pm 0.5$ \\
 MC  & $\ll H_\mu\rr $ (km)      & $41.6\pm0.3$ & $34.3\pm 0.3$ & $28.1\pm 0.3$ & $23.8\pm 0.3$ & $20.4\pm 0.3$ \\
     & $\ll E_p\rr $ (TeV)       & $4.1\pm0.2$ &  $4.0\pm 0.2$ & $ 4.0\pm 02$ &  $3.9\pm 0.1$ & $ 3.9\pm 0.2$ \\
     & $\ll P_\perp\rr $ (GeV/c) & $0.56\pm0.01$ & $0.59\pm 0.01$ &$0.57\pm 0.01$ &  $0.57\pm 0.02$ & $0.57\pm 0.01$ \\
\end{tabular}
\vskip 0.3cm
(a)
\vskip 0.7cm

0.8$<$cos$\theta$$<$0.9
\vskip 0.2cm
\begin{tabular}{c l c c c c c}
		      &
		      &
     3350$<$h$<$3750  &
     3750$<$h$<$4150  &
     4150$<$h$<$4550  &
     4550$<$h$<$4950  \\
		      &
		      &
     ($hg/cm^{2}$)    &
     ($hg/cm^{2}$)    &
     ($hg/cm^{2}$)    &
     ($hg/cm^{2}$)    \\
\tableline
 EXP & $\ll D\rr $ (m)           & $9.4\pm2.1$ & $8.5\pm1.9$ & $7.3\pm1.6$ & $6.2\pm1.6$ \\
\tableline
     & $\ll D\rr $ (m)           & $9.7\pm3.4$ & $8.8\pm1.2$ & $7.7\pm1.1$ & $7.1\pm1.1$ \\
     & $\ll X\rr $ (km)          & $42.7\pm 0.4$ & $42.5\pm .4$ & $45.9\pm 0.3$ & $43.76\pm 0.3$ \\
 MC  & $\ll H_\mu\rr $ (km)      & $23.7\pm 0.3$ & $23.8\pm 0.3$ & $24.6\pm 0.3$ & $25.1\pm 0.5$ \\
     & $\ll E_p\rr $ (TeV)       & $3.6\pm 0.1$ & $ 3.9\pm 0.1$ &  $4.4\pm 0.1$ & $ 4.8 \pm 0.02$ \\
     & $\ll P_\perp\rr $ (GeV/c) & $0.56\pm 0.02$ & $0.57\pm 0.02$ & $0.58\pm 0.02$ & $0.58\pm 0.02$ \\
\end{tabular}
\vskip 0.3cm
(b)
\end{center}
\caption{\em Average separation between muon pairs $\langle D \rangle$ 
(in m) as a function of cos$\theta$ (a) and rock depth (b).
In each table the experimental data are compared to the
expectations from the HEMAS Monte Carlo. For the same simulations, other
averages of relevant quantities are reported.
\label{t:tab0}}
\end{table}

\begin{table}[p]
\begin{center}
\begin{tabular}{l c c c c c}
& 0--80 cm & 80--160 cm & 160--240 cm & 240--320 cm & 320--400 cm \\ 
& & & & & (max) \\
\tableline
Exp. Data & 5528 & 12491 & 17569 & 20514 & 20816 \\
MC Data   & 5154 & 21417 & 33573 & 40367 & 42679 \\
Discrepancy after& (55$\pm$2)\% & (16$\pm$2)\% & (6$\pm$1)\% & (4$\pm$1)\% & \\ 
normalization & & & & & \\
\tableline
Exp. Data + C4 & 3612 & 11128 & 16535 & 19597 & 19977 \\
MC Data + C4& 4848 & 20346 & 31932 & 38425 & 40660 \\
Discrepancy after& (34$\pm$2)\% & (10$\pm$2)\% & (6$\pm$2)\% & (4$\pm$2)\% & \\ 
normalization & & & & & \\
\tableline
Exp. Data + C4 + & 2193 & 9264 & 15462 & 19190 & 19842 \\
$\mu$ pair subtraction & & & & & \\
MC Data + C4& 4848 & 20346 & 31932 & 38425 & 40660 \\
Discrepancy after& (8$\pm$7)\% & (7$\pm$3)\% & (0$\pm$2)\% &  (2$\pm$2)\% & \\ 
normalization & & & & & \\
\end{tabular} 
\end{center}
\caption{\em Number of weighted muon pairs in the first few bins of
the experimental and simulated decoherence distributions.
The discrepancy is the percentage difference between 
experimental and Monte Carlo values, normalized to the distribution maximum 
(last column).}
\label{t:tab1}
\end{table}

\end{document}